# Cloud Infrastructure Provenance Collection and Management to Reproduce Scientific Workflow Execution

Khawar Hasham[a,*], Kamran Munir[a,**], Richard McClatchey[a]

[a]*Centre for Complex Cooperative Systems (CCCS), Department of Computer Science and Creative Technologies (CSCT), University of the West of England (UWE), Frenchay Campus, Coldharbour Lane, Bristol, BS16 1QY, United Kingdom*

**Abstract**

The emergence of Cloud computing provides a new computing paradigm for scientific workflow execution. It provides dynamic, on-demand and scal- able resources that enable the processing of complex workflow-based experi- ments. With the ever growing size of the experimental data and increasingly complex processing workflows, the need for reproducibility has also become essential. Provenance has been thought of a mechanism to verify a workflow and to provide workflow reproducibility. One of the obstacles in reproduc- ing an experiment execution is the lack of information about the execution infrastructure in the collected provenance. This information becomes crit- ical in the context of Cloud in which resources are provisioned on-demand and by specifying resource configurations. Therefore, a mechanism is re- quired that enables capturing of infrastructure information along with the provenance of workflows executing on the Cloud to facilitate the re-creation of execution environment on the Cloud. This paper presents a framework, ReCAP, along with the proposed mapping approaches that aid in captur- ing the Cloud-aware provenance information and help in re-provisioning the execution resource on the Cloud with similar configurations. Experimental evaluation has shown the impact of different resource configurations on the workflow execution performance, therefore justifies the need for collecting such provenance information in the context of Cloud. The evaluation has also demonstrated that the proposed mapping approaches can capture Cloud

[*]Corresponding author
[**]Principle corresponding author
*Email addresses:* mian.ahmad@uwe.ac.uk (Khawar Hasham), kamran2.munir@uwe.ac.uk (Kamran Munir)

information in various Cloud usage scenarios without causing performance overhead and can also enable the re-provisioning of resources on Cloud. Experiments were conducted using workflows from different scientific domains such as astronomy and neuroscience to demonstrate the applicability of this research for different workflows.

## 1. Introduction

The scientific community is experiencing a data deluge due to the generation of large amounts of data in modern scientific experiments that include projects such as the Laser Interferometer Gravitational Wave Observatory (LIGO) (Abramovici et al., 1992), the Large Hadron Collider (LHC)[1], and projects such as neuGRID (Munir et al., 2014, 2015). In particular the neuGRID community is utilising scientific workflows to orchestrate the complex processing of its data analysis. A large pool of compute and data resources are required to process this data, which has been available through the Grid (Foster and Kesselman, 1999) and is now also being offered by the Cloud-based infrastructures.

Cloud computing (Mell and Grance, 2011) has emerged as a new computing and storage paradigm, which is dynamically scalable and usually works on a pay-as-you-go cost model. It aims to share resources to store data and to host services transparently among users at a massive scale (Mei et al., 2008). Its ability to provide an on-demand computing infrastructure with scalability enables distributed processing of complex scientific workflows for the scientific community (Deelman et al., 2008). (Juve and Deelman, 2010) has experimented with Cloud infrastructures to assess the feasibility of executing workflows on the Cloud.

An important consideration during this data processing is to gather data that can provide detailed information about both the input and the pro- cessed output data, and the processes involved to verify and repeat a work- flow execution. Such a data is termed as Provenance in the scientific litera- ture. Provenance is defined as the derivation history of an object (Simmhan et al., 2005). This information can be used to debug and verify the execu- tion of a workflow, to aid in error tracking and reproducibility. This is of vital importance for scientists in order to make their experiments verifiable and repeatable. This enables them to iterate on the scientific method, to evaluate the process and results of other experiments and to share their own experiments with other scientists (Azarnoosh et al., 2013). The execution of scientific workflows in Clouds brings to the fore the need to collect provenance information, which is necessary to ensure the reproducibility of these experiments.

---

[1]http://lhc.cern.ch

## 2. Motivation

A research study (Belhajjame et al., 2012) conducted to evaluate the reproducibility of scientific workflows has shown that around 80% of the workflows cannot be reproduced, and 12% of them are due to the lack of information about the execution environment (Santana-Pérez and Pérez-Hernández, 2014). This information affects a workflow execution in multiple ways. A workflow execution can not be reproduced if the underlying execution environment does not provide the libraries (i.e. software dependencies) that are required for workflow execution. Besides the software dependencies, hardware dependencies related to an execution environment can also affect a workflow execution. It can affect a workflows overall execution performance and also job failure rate. This effect on the experiment performance has also been highlighted by Kanwal et al. (2015). For instance, a data-intensive job can perform better with 2GB of RAM because it can accommodate more data in RAM, which is a faster medium than hard disk. However, the job's performance will degrade if a resource of 1GB RAM is allocated to this job as less data can be placed in RAM. Moreover, it is also possible that jobs will remain in waiting queues or fail during execution if their required hardware dependencies are not met. Therefore, it is important to collect the Cloud infrastructure or virtualization layer information along with the workflow provenance to recreate similar execution environment to ensure workflow reproducibility. However, capturing such an augmented prove- nance becomes more a challenging issue in the context of Cloud in which resources can be created or destroyed at runtime.

The Cloud computing presents a dynamic environment in which resources are provisioned on-demand. For this, a user submits resource configuration information as resource provision request to the Cloud infrastructure. A resource is allocated to the user if the Cloud infrastructure can meet the submitted resource configuration requirements. Moreover, the pay-as-you-go model in the Cloud puts constraints on the lifetime of a Cloud resource. For instance, one can acquire a resource for a lifetime but he has to pay for that much time. This means that a resource is released once a task is finished or payment has ceased. In order to acquire the same resource, one needs to know the configuration of that old resource. This is exactly the situation with repeating a workflow experiment on the Cloud. In order to repeat a workflow execution, a researcher should know the resource configurations used earlier in the Cloud. This enables him to re-provision similar resources and repeat workflow execution.

The dynamic and geographically distributed nature of Cloud computing makes the capturing and processing of provenance information a major research challenge (Zhao et al., 2011; Vouk, 2008). Contrary to Grid computing, the resources in the Cloud computing are virtualised and provisioned on-demand, and released when a task is complete (Foster et al., 2008). Generally, an execution in Cloud based environments occurs transparently to the scientist, i.e. the Cloud infrastructure behaves like a black box. Therefore, it is critical for scientists to know the parameters that have been used and what data products were generated in each execution of a given workflow (SMS et al., 2011; Shamdasani et al., 2012). Due to the dynamic nature of the Cloud the exact resource configuration should be known in order to reproduce the execution environment. Due to these reasons, there is a need to capture information about the Cloud infrastructure along with workflow provenance, to aid in the repeatability of experiments.

## 3. Related Work

Significant research (Foster et al., 2002; Scheidegger et al., 2008) has been carried out in workflow provenance for Grid-based workflow management systems. Chimera (Foster et al., 2002) is designed to manage the data-intensive analysis for high-energy physics (GriPhyN)[2] and astronomy (SDSS)(http://www.sdss.org) communities. It captures process information, which includes the runtime parameters, input data and the produced data. It stores this provenance information in its schema, which is based on a relational database. Although the schema allows storing the physical location of a machine, it does not support the hardware configuration and software environment in which a job was executed. VisTrails (Scheidegger et al., 2008) provides support for scientific data exploration and visualization. It not only captures the execution log of a workflow but also the changes a user makes to refine his workflow. However, it does not support the Cloud vir-tualization layer information. Similar is the case with Pegasus/Wings (Kim et al., 2008) that supports evolution of a workflow. However, this paper is focusing on the workflow execution provenance on the Cloud, rather than the provenance of a workflow itself (e.g. design changes).

---

[2]http://www.phys.utb.edu/griphyn/

There have been a few research studies (e.g. de Oliveira et al. (2010); Ko et al. (2011)) performed to capture provenance in the Cloud. However, they lack the support for workflow reproducibility. Some of the work in Cloud towards provenance is directed to the file system (Zhang et al., 2011; Tan et al., 2012) or hypervisor level (Macko et al., 2011). However, such work is not relatable to our approach because this paper focuses on vir- tualized layer information of the Cloud for workflow execution. Moreover, the collected provenance data provides information about the file access but it does not provide information about the resource configuration. The PRECIP (Azarnoosh et al., 2013) project provides an API to provision and execute workflows. However, it does not provide provenance information of a workflow.

There have been a few recent projects (e.g. Chirigati et al. (2013); Janin et al. (2014)) and research studies e.g. (Santana-Perez et al., 2014a) on collecting provenance and using it to reproduce an experiment. A semantic-based approach (Santana-Perez et al., 2014a) has been proposed to improve reproducibility of workflows in the Cloud. This approach uses ontologies to extract information about the computational environment from the annotations provided by a user. This information is then used to recreate (install or configure) that environment to reproduce a workflow execution. On the contrary, our approach is not relying on annotations rather it directly interacts with the Cloud middleware at runtime to acquire resource configuration information and then establishes mapping between workflow jobs and Cloud resources. The ReproZip software (Chirigati et al., 2013) uses system call traces to provide provenance information for job reproducibility and portability. It can capture and organize files/libraries used by a job. The collected information along with all the used system files are zipped together for portability and reproducibility purposes. Similarly, a Linux-based tool, CARE (Janin et al., 2014), is designed to reproduce a job execution. It builds an archive that contains selected executable/binaries and files accessed by a given job during an observation run. Both these approach are useful at individual job level but are not applicable to an entire workflow, which is the focus of this paper. Moreover, they do not maintain the hardware configuration of the underlined execution machine. Furthermore, these approaches operate along with the job on the virtual machine. On the contrary, out proposed approach works outside the virtual machine and therefore does not interfere with job execution.

## 4. Workflow Reproducibility Requirements for Cloud

According to the current understanding of available literature, there is not a standard reproducibility model proposed thus far for scientific work- flows, especially in a Cloud environment. However, there are some guidelines or policies, which have been highlighted in the literature to reproduce experiments. There has been one important effort by C. (2010) in this regard, but this mainly talks about reproducible papers and it does not consider the execution environment of workflows. The same concern has been shared by Santana-Perez et al. (2014b) that most of the approaches in the conserva- tion of computational science, in particular for scientific workflow executions, have been focused on data, code, and the workflow description. They do not focus on the underlying infrastructure, which is composed of a set of computational resources (e.g. execution nodes, storage devices, networking) and software components. A recent study (Banati et al., 2015) emphasised on the need of incorporating the infrastructure information in the collected provenance. In this section, a few basic points are gathered from litera- ture analysis and Cloud context to present a set of workflow reproducibility requirements in the Cloud. These points also provide the basis for the pro- posed solution for workflow execution reproducibility on the Cloud. These points are discussed as follows.

(i) **Code and Data Sharing**
    The need for data and code sharing in computational science has been widely discussed (C., 2010). Code must be available to be distributed, and data must be accessible in a readable format (Santana-Perez et al., 2014a). In computational science, particularly for scientific workflow executions, it is emphasized that the data, code, and the workflow description should be available in order to reproduce an experiment. In the absence of such information or data, experiment reproducibility cannot be achieved because different results would be produced if the input data changes. It is also possible that the experiment cannot be successfully executed in the absence of the required code and its dependencies.
(ii) **Execution Infrastructure**
    A workflow is executed on an infrastructure provided by the Grid or the Cloud. The execution infrastructure is composed of a set of com- putational resources (e.g. execution nodes, storage devices, network- ing). The physical approach, where actual computational hardware

are made available for long time periods to scientists, often conserves the computational environment including supercomputers, clusters, or Grids (Santana-Perez et al., 2014b). As a result, scientists are able to reproduce their experiments in the same hardware environment. How- ever, this luxury is not available in the Cloud in which resources are virtual and dynamic. Therefore, it is important to collect the Cloud re- source information in such a manner that will assist in re-provisioning of similar resources on the Cloud for workflow re-execution. This will enable a researcher to recreate a virtual machine with similar resource configurations. Banati et al. (2015) also emphasized on the need to incorporate infrastructure information as part of the workflow prove- nance.

From a resource provisioning as well as a performance point of view, the following factors are important in selecting appropriate resources especially on the Cloud. These factors include: RAM, vCPU, Hard Disk, CPU Speed in MIPS. All these factors contribute to the job's execution performance as well as to its failure rate. For instance, con- sider a job that requires 2 GB of RAM during its processing. This job will fail if it is scheduled to a resource with less available RAM. Moreover, it could also affect its performance if more and more data is processed from hard disk. Similarly, vCPU (virtual CPUs, meaning CPU cores) along with the MIPS value directly affect the job exe- cution performance. In a study (Vo¨ckler et al., 2011), it was found that the workflow task durations differ for each major Cloud, despite the identical setup. It was suggested that lower/different CPU speed, and a poor WAN performance could be one factor for different or slow workflow execution times.

Hard disk capacity also becomes an important factor in provisioning a new resource on the Cloud. It was argued that building images for scientific applications requires adequate storage within a virtual machine. In addition to the OS and the application software, this storage is used to hold job inputs and output that are consumed and produced by a workflow job executing on the VM (Vo¨ckler et al., 2011). Out of these factors, current Cloud offerings only support the provi- sion of resources based on RAM, vCPU and Hard Disk. These factors are combined and named as instance type (e.g.    in Amazon EC2[3]), or flavour (e.g. in OpenStack[4]). The MIPS information is not provided as a parameter for acquiring a resource. Therefore, the proposed architecture in this paper takes these three factors along with the software environment (discussed below) into consideration for resource provisioning. Nonetheless, the efficacy of having MIPS information in the collected provenance will be shown through our results (discussed in Section 8). This will aid in providing a motivation and envisioning a future possibility in which the Cloud Providers will start this as a configurable parameter of a resource.

---

[3]http://aws.amazon.com/ec2

(iii) **Software Environment**

Apart from knowing the hardware infrastructure, it is also essential to provide information about the software environment. A software environment determines the operating system and the libraries used to execute a job. Without the access to required library information, a job execution will fail. For example, a job, relying on a MATLAB library, will fail in the case where the required library is missing. One possible approach (Howe, 2012) to conserve software environments is thought to conserve the VM that is used to execute a job and then reuse the same VM while re-executing the same job. One possible mechanism is to create snapshot of virtual machines for each job, however the high storage demand of VM images poses a challenging problem (Zhao et al., 2014). In the prototype proposed in this research study, the VM is assumed to present all the software dependencies required for a job execution in a workflow. Therefore, the proposed solution will also retrieve the image information in building a virtual machine on which the workflow job was executed.

(iv) **Workflow Versioning**

Scientific workflows are commonly subject to a reduced ability to be executed or repeated, largely due to the volatility of the external resources that are required for their executions (Gómez-Pérez et al., 2013). Capturing only a provenance trace is not sufficient to allow the computation to be repeated a situation known as workflow decay (Roure et al., 2011). The reason is that the provenance systems can store information on how the data was generated, however they do not store copies of the key actors in the computation i.e. workflow, services, data. Workflow versioning along with other provenance information has been suggested to achieve reproducibility (Woodman et al., 2011). Recently, Sandve et al. (2013) have suggested archiving the exact versions of all programs and enabling version control on all scripts used in an experiment. This is not supported in the presented prototype because the focus of this research study is on the execution aspect of a workflow. Nonetheless, it can be incorporated in future by using CRISTAL since it can track the evolution of its stored items (Branson et al., 2012). Since the focus of this research work is on the workflow execution phase, this aspect has consequently not been dis- cussed in detail, however, the original workflow description along with its associated files has been stored to support workflow reproducibility of the same workflow.

---

[4]http://openstack.org

(v) **Provenance Comparison**

The provenance of workflows should be compared to determine workflow reproducibility. The comparison should be made at different levels; workflow structure, execution infrastructure, and workflow input and output. A brief description of this comparison is given below:

(a) Workflow structure should be compared to determine that both workflows are similar. Because it is possible that two workflows may have a similar number of jobs but with a different job execution order.

(b) Execution infrastructure (i.e. the software environment and resource configuration) used for a workflow execution should also be compared.

(c) Comparison of the inputs and outputs should be made to confirm workflow reproducibility. There could be a scenario in which a user repeated a workflow but with different inputs, thus producing different outputs. It is also possible that changes in job or software library results into a different workflow output.

In general, the provenance approach used by most Workflow Management Systems (WMS) such as Kepler or VisTrail enforces this strict reproducibility requirement for relatively small amounts of consumed and produced data, or just for primitive data. The most common strategy is to save all consumed and produced data into a relational database, together with a link among data and the corresponding execution (Lifschitz et al., 2011). However, this approach is not feasible for large data files.

There are a few approaches (e.g. Missier et al. (2014)) that perform a comparison on workflow provenance graphs to determine differences in reproduced workflows. The proposed approach in this research study incorporates the workflow structure and infrastructure along with output comparison to determine the reproducibility of a workflow. The important difference from Missier's approach is the comparison of Cloud infrastructure information in the provenance graph. Since this paper is focusing on collecting Cloud resource information during workflow execution, the execution infrastructure comparison has been used to evaluate the proposed approaches (see Section 8).

## 5. Workflow Execution Scenario on the Cloud

There have been projects e.g. (CERNVM, 2016) that uses the Cloud for the execution of workflows. Mainly, these approaches create a virtual environment i.e. a virtual Grid on top of the Cloud resources using their legacy systems and execute workflows. A system, AMOS (Strijkers et al., 2010), presents a layer on top of a workflow management system and dynamically creates resources on the Cloud to instantiate a transient Grid ready for immediate use in the Cloud. A similar approach has also been discussed and tested by (Vöckler et al., 2011; Juve et al., 2009). It uses Pegasus as a WMS along with the Condor (Tannenbaum et al., 2002) cluster on the Cloud infrastructure to execute workflow jobs. In this section, a scenario (see Figure 1) is presented that can be used to execute a workflow on the Cloud.

A scientist creates a workflow using a workflow authoring tool or uses an existing workflow from the Pegasus Provenance Store e.g. database and submits it to the Cloud infrastructure through Pegasus. Pegasus interacts with a cluster of compute resources in the form of Condor instances running on virtual machines (VM) in the Cloud. Each VM has a Condor instance to execute the users job.

Pegasus schedules the workflow jobs to these Condor instances and retrieves the workflow provenance information supported by the Pegasus database. The collected provenance information, which is stored in the Pegasus database, comprises job arguments (input and outputs), job logs (output and error) and host information. However, the collected host information is not sufficient to re-provision resources on the Cloud because Pegasus was designed initially for the Grid environment, and such systems lack this capability at the moment (as discussed in Section 3). This workflow execution scenario on Cloud has inspired the architecture of the proposed system presented in (Hasham et al., 2014). The following Section 6 discusses the proposed architecture of ReCAP in detail. It captures the Cloud resource information and links it with the workflow provenance to generate Cloud-aware provenance (CAP).

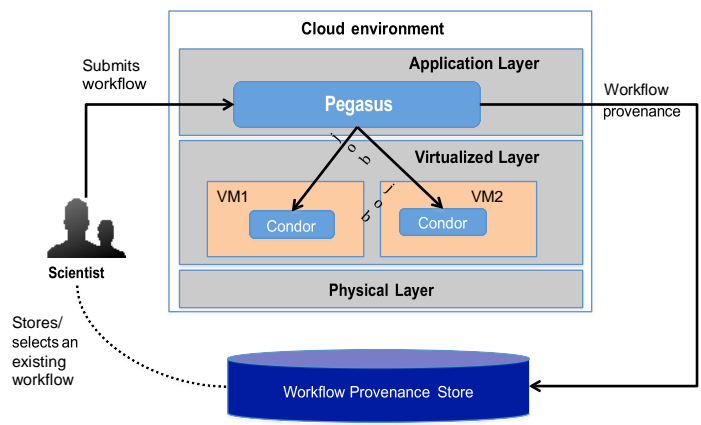

**Figure 1:** Workflow Execution on Cloud

## 6. ReCAP: Reproduce Workflow execution using Cloud-Aware Provenance

This section presents the detailed architecture of the proposed system, ReCAP, that has been designed on the configuration and plugin-based mechanism. With this mechanism, support for new workflow management systems, mapping algorithms etc. can be easily added without changing the core of the system. There are seven key components in this design. They are the (i) WMS Wrapper Service, (ii) WS Client, (iii) WMS Layer, (iv) Cloud Layer, (v) Aggregator, (vi) WF-Repeat, and (vii) Comparator. Each of these components can further have their sub-components which are also discussed in this section. Figure 2 shows the detailed architecture of ReCAP and mutual interaction among its components.

### 6.1. ReCAP Configuration

ReCAP is designed using a plugin based approach and this requires a set of configuration parameters to drive the overall system. Consequently, the key aspects of the ReCAP such as WMS components, mapping algorithms, persistence API that interacts with the workflow provenance, the ReCAP databases and the Cloud middleware are driven by the configuration parameters. These configurations (shown as *ReCAP configs* in Figure 2) are

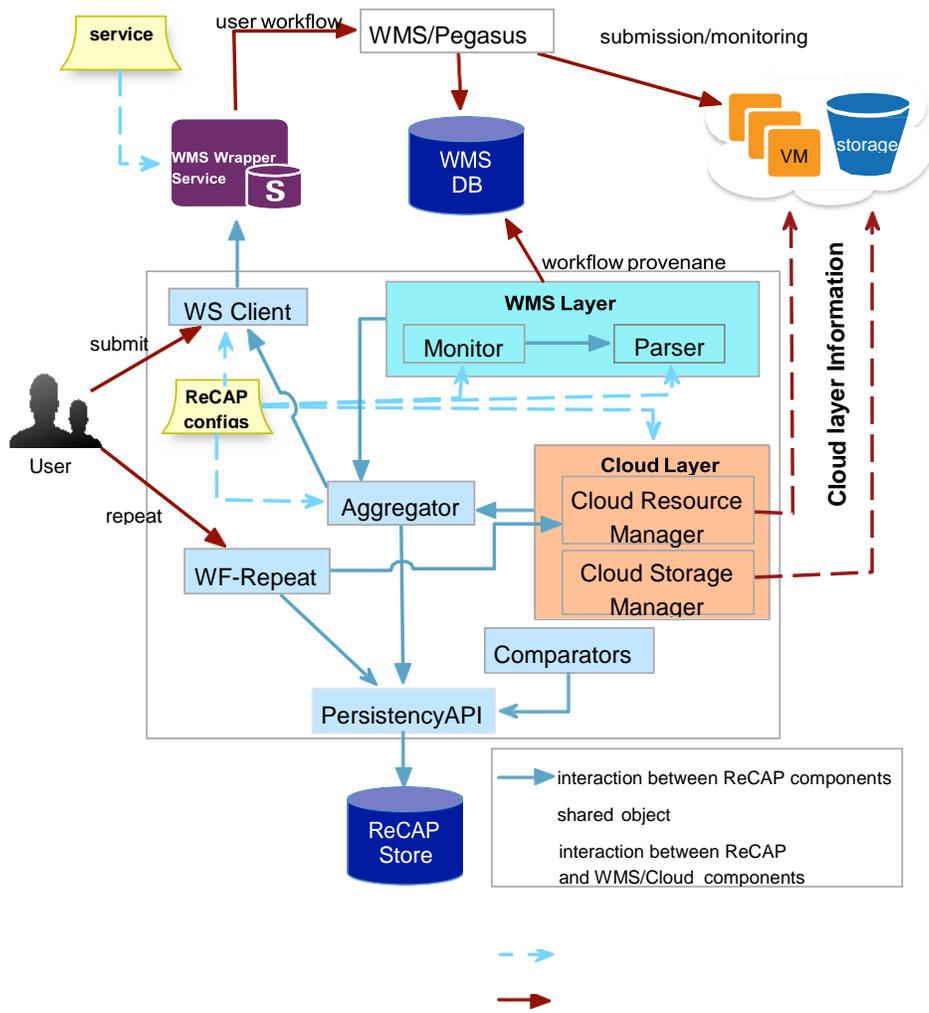

**Figure 2:** Detailed architecture of the ReCAP system

divided into seven main sections as shown in Listing 6.1. These sections are discussed as follow:

- **cloud settings**: Provides a large number of parameters which are mainly used to access the Cloud middleware. The implemented classes accessing the Cloud middleware to retrieve a Cloud resource, a Virtual machine, information establish connection with the Cloud middleware using these configured parameters. Since the ReCAP prototype is using Apache Libcloud[5] API, which can interact with various Cloud middlewares, these parameters can be changed to accommodate a new Cloud middleware without any change in the code. Another impor- tant parameter in this section is *MAPPING TYPE*, which informs the ReCAP to load the appropriate mapping algorithm.

- **storage settings**: Provides access parameters to access storage ser- vice on the Cloud. Using these parameters, the API establishes con- nection with the Cloud storage service. Since the ReCAP prototype is using Apache Libcloud API, which can interact with various Cloud middlewares, these parameters can be changed to accommodate a new Cloud middleware without any change in the code.

- **wmsdb settings**: Provides information about the parameters used to connect with the database of the workflow management system. Since the prototype's persistency layer is built on top of an SQLAlchemy framework, which provides access to many databases such as Oracle, Sqlite, MSSQL etc., changing a database would not require a change in the persistency layer. For this prototype, Pegasus database settings on MySQL database have been used.

- **recapdb settings**: Provides information about the parameters used to connect with the database used in the prototype. This database contains the relational schema shown in Section 6.10 and holds the mapping information between a job and a Cloud resource.

- **WMS settings**: Depending upon the used workflow management system, these parameters can be changed. For instance, *wms monitor* parameter loads the appropriate monitoring component that monitors the workflow state in the database configured in above settings. This component is WMS specific and so is its implementation.

---

[5]https://libcloud.apache.org/

- **WrapperService**: To interact with the WrapperService, the Client component requires connection information such as service URL and user credentials. This section provides these details.

- **log settings**: In order to log the inner activities of ReCAP, logging is provided and it is controlled by this parameter.

```
[cloud_settings]
swift_host=164.11.100.72
service_name=Compute Service
OS_USERNAME=xxxxxxxx
OS_PASSWORD=xxxxxxxx
OS_AUTH_URL=https://api.opensciencedatacloud.org:5000/sullivan/v2.0/tokens
OS_TENANT_NAME=xxxxxxxxx
OS_REGION_NAME=RegionOne
#mapping types could be static,eager,lazy
MAPPING_TYPE=static
[storage_settings]
swift_host=<SERVER_IP>
OS_USERNAME=xxxxxxxxx
OS_PASSWORD=xxxxxxxxx
OS_AUTH_URL=http://<SERVER_IP>:5000
OS_TENANT_NAME=admin
OS_REGION_NAME=UWE_Region
EC2_URL=http://<SERVER_IP>:8773/services/Cloud
EC2_ACCESS_KEY=<EC2_ACCESS_KEY>
EC2_SECRET_KEY=<EC2_SECRET_KEY>
[wmsdb_settings]
user=pegUser
password=xxxxxxx
host=<DBSERVER_IP>
port=3306
dburl=mysql+mysqlconnector://pegUser:xxxxxxx@<DBSERVER_IP>/pegasusdb
database=pegasusdb
[recapdb_settings]
user=CAPuser
password=xxxxxxxx
host=<DBSERVER_IP>
port=3306
dburl=mysql+mysqlconnector://CAPuser:xxxxxxx@<DBSERVER_IP>/recapdb
database=recapdb
[WMS_settings]
wms_monitor=PegasusMonitor
wms_parser=PegasusParser
```

```
[WrapperService]
endpoint=http://<SERVER_IP>:5000/service_wrapper/api/v1.0
service_user=khawar
service_password=XXXXX
[log_settings]
log_conf=/opt/MultiLayerProv/conf/logging.conf
```

**Listing 1:** ReCAP prototype configurations

### 6.2. WMS Wrapper Service

This component of ReCAP is a RESTful web service that operates on top of a workflow management system, which for this study is Pegasus. It exposes interfaces through which a user can interact with the underlining workflow management system and can submit his workflows. As this service mainly interacts with a WMS, it resides on the same machine on which that system is running. For instance, the machine on which Pegasus runs and uses to submit workflows is called the Submit Host. In our case, Pegasus uses the Condor pool to execute workflow jobs. In this environment, the *SubmitHost* is configured as the Master node for the Condor pool and all the VMs with the Condor instances acting as worker nodes. All these nodes create a Condor pool over the virtual machines which is termed a *V irtualCluster* Juve and Deelman (2010). Figure 3 illustrates the interaction of the WMS with other components of the system. In the case of Pegasus and Condor, as shown in Figure 3, this service interacts with Pegasus to submit the workflows received in the user's request. It also interacts if needed with the Condor pool through the *SubmitHost*. This interaction is useful for achieving Eager Resource-Job mapping (will be discussed later in 7.2).

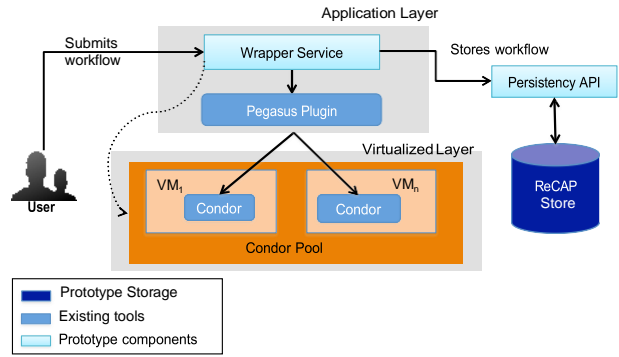

**Figure 3:** Interaction of Service Wrapper with overall system or its architecture

A user accesses the Wrapper Services over HTTP and requests to submit his workflow. In this request, he will provide the abstract representation of his workflow i.e. DAX and its associated configurations, which are specific to the underlining workflow management system. In the case of Pegasus, he will provide a DAX file representing his workflow and a site file that provides information about storage, environment variables and workflow constraints. Upon receiving the request, the Service Wrapper firstly authenticates the user with the provided credentials in the service request. This enables the service to prevent unauthorized access to the underlying resources. At the moment, a very basic password-based HTTP authentication is implemented. However, it can be extended to database driven or more complex user au- thentication model.

After successful user authentication, the Service Wrapper loads the appropriate WMS plugin - specified in the service configuration - that can interact with the underlining workflow management system. As this study is using Pegasus, a Pegasus plugin is implemented that interacts with Pegasus commands and its database. The plugin stores the user provided files locally. The storage location on the local file system is controlled by a configuration parameter. After storing files, it then submits them to Pegasus using *pegasus − plan*[6]. This command parses the given abstract workflow into concrete workflow (DAG) and submits it to the Condor pool using Condor's DAGMan, which is responsible for managing DAG representation. An executable workflow, with all jobs and their dependencies, both control and data, is represented in the DAG representation. Once the workflow has submitted, Pegasus provides a unique identifier for this workflow and that is returned back to the user or a component as a response to its workflow submission request. This response is returned in JSON format, which is compact, flexible (schemaless) and easier to use. Upon successful submission of the user's workflow, the Service Wrapper stores the provided files in the database for later use in reproducing a workflow execution. The Figure 4 illustrates the flow of activities within the Service Wrapper for submitting a user workflow.

The following main operations are supported at the time of writing, which can be called from the service Client component.

- **submit**: This operation is used to submit a workflow to the underlying workflow management system, which is Pegasus in this prototype. All required files are passed in the service request, which are processed accordingly.

---

[6]pegasus-plan command http://pegasus.isi.edu/mapper/docs/4.0/cli-pegasus-plan.php

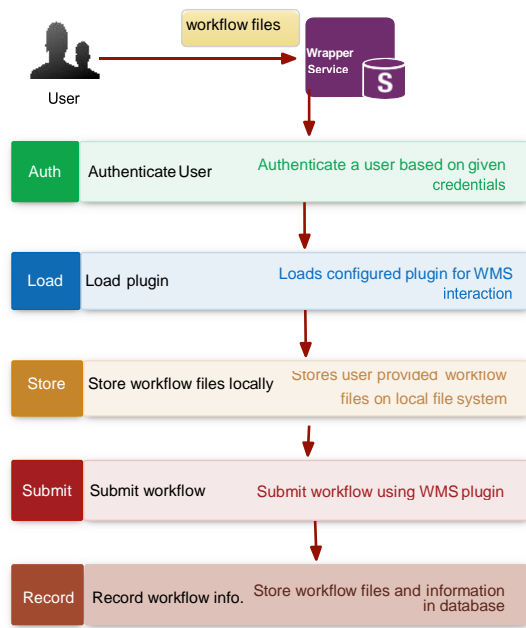

**Figure 4:** Flow chart of WMS Wrapper Service to process a workflow submission request

- **wms_get_file**: This operation is used to retrieve files (job outputs, workflow submission output) stored within the workflow management system's work directory. It takes two arguments i.e. (1) job infor- mation and (2) the directory location to find the requested files. This operation is used while processing the job logs in mapping components.

- **jobmon**: This operation is Condor-specific as it attempts to retrieve the current status of the job running on Condor. It helps in retrieving the host information from the Condor pool on which a job is running. This operation is used in the Eager approach (discussed later in Section 7.2).

- **cpool_mips**: This operation is used to to retrieve the MIPS of the machines in the Condor pool. MIPS or KFLOPS are one way to specify the execution performance of a machine and it can affect a job execution performance (shown and discussed later in Section 8).

### 6.3. WS Client

In order to interact with the Wrapper Service, the WS Client component of ReCAP is used. All interactions with the Wrapper Service pass through the WS Client component. On receiving requests from a user or other components such as the Monitor component of ReCAP, the Client component starts an HTTP session with the WrapperService. As authentication has been implemented in the WrapperService to avoid malicious access, it also provides user credentials along with the request. These settings are re- trieved from the ReCAP configurations (see *WrapperService* in Listing 6.1) discussed earlier in Section 6.1.

In order to submit a workflow, the user interacts with the Client compo- nent and passes all required files. The Client component interacts with the WrapperService and submits the files. It retrieves the response from the WrapperService and uses the configured WMS Parser to parse it to extract the workflow ID and ReCAP ID assigned to it. It then starts the Monitor component to start monitoring the provenance information of the submitted workflow. This interaction is shown in Figure 5.

### 6.4. WMS Layer

As discussed earlier, the design philosophy of ReCAP is plugin-based in or- der to provide support for extensibility. A literature review has shown that multiple workflow management systems such as Pegasus, Chimera, Taverna and Kepler etc., provide either the hostname or the IP of the machine on which a job was executed. Therefore, it has been considered to enable support for multiple workflow management systems by adopting a plugin-based design. This section provides detail about the WMS Monitor and Parser components. During mapping process, this layer loads appropriate plugin implementation based upon the configuration parameters (see WMS settings section in Listing 6.1). In this prototype, Pegasus-based plugins have been developed and tested.

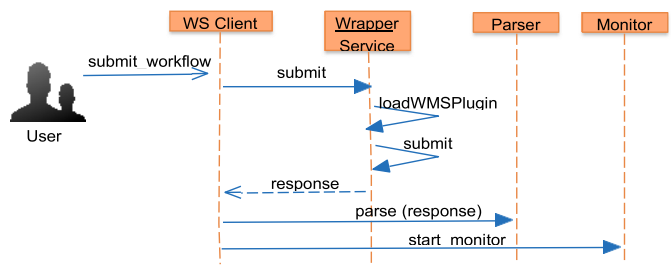

**Figure 5:** Illustrating the interaction between ReCAP components for submitting a user workflow

### 6.4.1. Monitor

This component has been designed to monitor a workflow execution and retrieving its provenance information. This component presents a threaded implementation which enables continuous monitoring of a workflow execution. In order to perform monitoring operations, it interacts with the WMS database and retrieves workflow and job states. Once a workflow is finished, it starts the job to Cloud resource mapping operation using the configured mapper plugin. It also interacts with the Parser plugin to parse job outputs. A monitor plugin for Pegasus has been written in this prototype.

### 6.4.2. Parser

This component helps in parsing the files and job outputs produced during a workflow execution. Since each WMS can produce its own files during workflow execution and the job output formats can also be WMS specific, this component helps in providing an abstraction layer on top. For example, once a workflow is submitted through Pegasus, a submit file is produced that contains information about the output directory and workflow identifiers. Moreover, job output logs contains information about the location of the input and output files on the Cloud. These parsers can also be extended to extract the CPU spec from the job outputs if they cannot be retrieved from the workflow provenance database. All this information is important to feed the Monitor component and also to acquire provenance information about the consumed or produced output files on the Cloud, which is later used in workflow output comparison algorithm (discussed Hasham et al. (2014)). ReCAP loads the Parser plugin using the global configuration. For instance, to parse Pegasus outputs, it loads the PegasusParser plugin.

### 6.5. Cloud Layer

To interact with the Cloud middleware, a component Cloud Layer named 'CloudLayerComponent' has been developed. It provides two types of interactions with the Cloud i.e. (a) to retrieve information about the virtual machines and also (b) to retrieve information about the workflow's input and output files stored on the Cloud storage service. These two types of interactions are handled by its two sub-components: the CRM and the CSM. These two components are briefly discussed in following sections.

### 6.5.1. Cloud Resource Manager (CRM)

The CRM interacts with the Cloud IaaS service such as the *nova* service of OpenStack to manage virtual resources on the Cloud. The management involves operations such as retrieving resource information of virtual machines and the provisioning of new resources on the Cloud upon receiving new resource requests. The retrieved information about the currently run- ning virtual machines include their metadata, OS images used in those VMs, flavours configuration used in VMs. This interaction is shown in Figure 6.

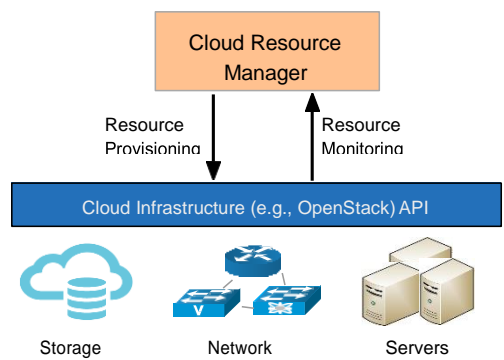

**Figure 6:** Interaction of CRM with the underlying Cloud infrastructure through APIs

### 6.5.2. Cloud Storage Manager (CSM)

This component interacts with the Cloud storage service such as Swift for the OpenStack Cloud middleware. As discussed earlier, the execution environment uses the Cloud storage service to save workflow inputs and outputs. Through this component, ReCAP is able to retrieve the files and their metadata from the Cloud. While storing the Cloud-aware provenance in the database, the Aggregator component invokes this component to retrieve filenames and their metadata such as MD5 hash and creation time etc. from a given location on the Cloud storage service. This component is also used to iterate over the produced files on the Cloud during the workflow output comparison (discussed in Section Hasham et al. (2014)).

### 6.6. Aggregator

The Aggregator (or also named Provenance Aggregator) component performs the mapping between the workflow job information collected from the Workflow Provenance component and the cloud resource information collected from the Cloud Layer Provenance component. In order to establish a mapping, it loads an appropriate mapping algorithm and this is driven by a configuration parameter *MAPPING TYPE* (as discussed above). The mapping information is then stored in the database. Section 7 explains the mapping algorithms designed in this prototype.

### 6.7. WF-Repeat

This component is designed to re-execute a workflow on the Cloud. A user can select a previous workflow, identified with a unique ID, and request this component to re-execute it. Upon receiving the request, this component retrieves the required workflow source files from the ReCAP database and also Cloud-aware provenance information. Using this Cloud-aware provenance information, it re-provisions the resources on the Cloud infrastructure and then submits the workflow over them using the underlying workflow management system, which is Pegasus in this research.

### 6.8. Comparator

The comparator component performs various provenance comparison operations for evaluating the workflow reproducibility. The comparisons include a workflow output comparison and a workflow graph structure comparison etc. It also performs provenance completeness and correctness analysis on the given workflow provenance traces. It takes two workflow identifiers to retrieve their provenance information from the database and then invokes the appropriate comparator implementation such as workflow output comparison.

### 6.9. Persistency API

This acts as a thin layer to expose the provenance storage capabilities to other components. In order to interact with the underlying databases, a persistency layer is designed to provide a common callable interface for mul- tiple backend engine such as MySQL[7]. This API uses SQLAlchemy[8] that provides access to multiple database engines. An instance to the database connection can be obtained by specifying the connection parameters in the config file (see *wmsdb settings* and *recapdb settings* in Listing 6.1).

### 6.10. ReCAP database schema

A relational database schema has been designed that assists in storing workflow descriptions and their configuration files, workflow job-to-Cloud resource mappings, temporary mappings for the *Lazy* approach or the *SNoHi* approach, and files metadata (consumed or produced data by the workflow jobs) stored on the Cloud. This information later helps in retrieving CAP information and also helps in answering Cloud-aware provenance queries and comparing workflow outputs. Following paragraphs describe the schema shown in Figure 7.

In order to preserve the original workflow and its associated configuration files, the *WorkflowSource* table is used. The *wfDAG* column stores the workflow representation described in a DAG format. This is an abstract workflow which will be submitted through Pegasus for execution. The *wfSite* column stores the information related to execution site and its storage elements. This information specifies the storage locations and paths to be used for reading and writing data during the workflow execution. The *wfTC* column stores information about the executables which are to be used for workflow jobs. A user can also provide configuration properties to Pegasus that affects the way Pegasus plans, schedules and stores workflows. This set of properties is stored in the *wfProps* column. Each workflow is assigned a unique ID by Pegasus and also in the ReCAP database, thus a mapping between these two IDs are required. The *wfID* is the ID assigned to a workflow execution by ReCAP and the *wms_wfid* is the unique ID assigned to a workflow by the workflow management system such as Pegasus. The *wms_wfid* is essential to record because it is used in retrieving workflow provenance from the Pegasus. Since the prototype is designed using the Pegasus WMS, this table contains a few Pegasus specific entries. However, the same concepts can also be applied when used with other similar workflow management systems.

---

[7]https://www.mysql.com/
[8]http://www.sqlalchemy.org/

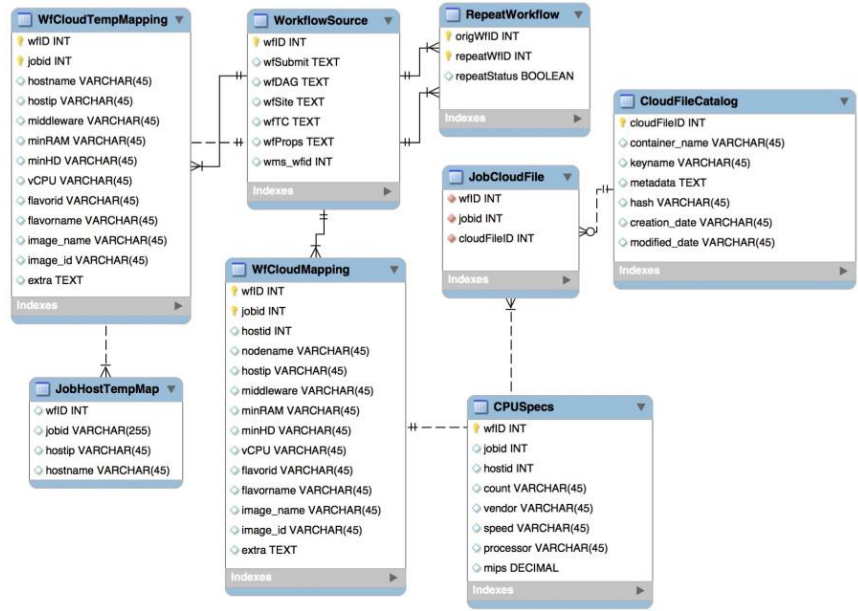

**Figure 7:** ReCAP Relational Database Schema

In order to establish the final job-to-Cloud resource mapping and to record Cloud-aware provenance information, the *WfCloudMapping* table is used. This table stores the mapping between workflow jobs and the configurations of the Cloud resources used for their execution. In order to specify the resource flavour, that provides the resource hardware configurations, *flavor- name* and *flavorid* are stored. However, this information can be customised or new entries can be added by the Cloud Provider. For instance, Amazon EC2 provides an extensive list[9] of instance types with customized values. On the other hand, Openstack middleware used in Open Science Data Cloud (OSDC)[10] provides a limited set of instance types. Therefore, the individual parameters i.e. *minRAM*, *minHD*, *minCPU* specifying a Cloud resource are also stored. As highlighted in Section 3 OR 4, the information about the software stack is also essential to successfully conduct an execution. The *image name* column provides the information about the operating system running the virtual machine. In this prototype, it is assumed that the operating system or the image contains all the required libraries on which a job executable is dependent. In order to store the billing information, which is also identified in Section 4, the *extra* column is stored provided the Cloud APIs support this functionality. This column basically stores data in a JSON format and thus enables us to store multiple key-value pairs in it. Due to this, it is also possible to store metadata information of a Cloud resource i.e. creation date, hash, etc. provided by the Cloud provider. This column has also been used to store the cost associated to the Cloud resource.

---

[9]http://aws.amazon.com/ec2/instance-types/
[10]https://www.opensciencedatacloud.org/

In order to tackle the dynamic resource scenario on the Cloud (discussed later in Section 7), the *WfCloudTempMapping* table is used. This table stores temporary mapping information which is then moved to the *WfCloudMapping* table once a workflow execution has finished. There is another table, *JobHostTempMap*, that stores temporary job and its execution host mapping. This table is used in the SNoHi mapping approach (discussed later in Section 7.4). In order to store more detailed information about the CPU, the *CPUSpecs* table is used. Although, the existing Cloud providers and their offerings do not allow a user to request a resource with such re- source parameters, the impact of CPU performance still cannot be ignored especially for compute-intensive jobs. Moreover, this information is also helpful when sharing experimental setup and results with the peers. This is why this information is captured and stored as part of the Cloud-aware provenance.

As discussed earlier, the data files are stored on the Cloud, this is why it is important to keep track of file locations and metadata on the Cloud. To achieve this, the *JobCloudFile* and *CloudFileCatalog* tables are conceived. The *JobCloudFile* table stores the mapping between workflow jobs and its produced/consumed files on the Cloud. The *CloudFileCatalog* provides detailed information about a file stored on the Cloud. This information in- cludes the location of the file, specified by the *container name* and *keyname*, MD5 *hash* of the file contents, additional *metadata* stored along with the file, *creation date* and *modified date*. By using this information, it is not only able to provide information about the file, but also can be helpful in comparing the file contents produced from workflow repeated executions to verify a workflow result. This also helps in identifying if a file is changed over time, which, however, is not the focus of this research study. For in- stance, a file created by a job will have the same creation and modification time or even no modification time. However, if a file is modified or tampered with somehow, the modification date will be updated. By comparing the latest dates with the already stored information, one can deduce if the file contents are changed.

## 7. Job-to-Cloud Resource Mapping

In order to reproduce workflow execution on the Cloud infrastructure, it is important to first collect such information as part of workflow execution provenance. This section discusses the job-to-Cloud resource mapping approaches and Cloud resource information, which is later used for re-executing a workflow on similar Cloud resources. Before diving into a detailed discussion of these approaches, first it is important to understand two different resource usage scenarios on the Cloud. These scenarios and their understanding provide a better picture of the requirements and the motivation behind devising different approaches to establish a job-to-Cloud resource mapping for each discussed scenario.

- **Static Environment on Cloud**

    In this environment, the virtual resources, once provisioned, remain in a *RUNNING* state on the Cloud for a longer time. This means that the resources will be accessible even after a workflow's execution is finished. This environment is similar to creating a virtual pool or Grid on top of Cloud's resources. A Static mapping scheme devised for such an environment will be discussed in Section 7.1

- **Dynamic Environment on Cloud**

    In this environment, VMs are shut down after the job is done. Therefore, a virtual resource, which was used to execute a job, will not be accessible once a job has finished. Moreover, in this environment, resources are provisioned on demand and released when they are no more required. Two approaches (a) Eager and (b) Lazy have been devised to handle this scenario and they will be discussed in Sections 7.2 and 7.3 respectively.

The mapping approaches discussed in the following sections achieve the job-to-Cloud resource mapping using the existing provenance information, which is available in many workflow management systems such as Pegasus or Chiron. One such information is an indication of the execution host or its IP in the collected provenance information, which is available across almost all existing workflow management systems. Many systems do maintain either name or IP information. In the Cloud's IaaS layer across one provider or for one user, no two machines can have same name or same IP at any given time. This means any running virtual machines should have unique IP. Once this virtual machine is destroyed, it is possible that the same IP is assigned later to a new virtual machine. All the rest of the properties of a virtual machine are accessible through IaaS layer and can be used by multiple ma- chines at a time. For instance, multiple machines can be provisioned with flavour *m1.small* or with OS image Ububtu 14.04.

*7.1. Static Approach*

The CloudLayerProvenance component of ReCAP is designed in such a way that interacts with the Cloud infrastructure as an outside client to obtain the resource configuration information. As mentioned earlier in Section 7, this information is later used for re-provisioning the resources to provide a similar execution infrastructure in order to repeat a workflow execution. Once a workflow has been executed, Pegasus collects the provenance and stores it in its own internal database. Pegasus also stores the IP address of the virtual machine (VM) where the job is executed. However, it lacks other VM specifications such as RAM, CPUs, hard disk etc. The Provenance Aggregator component retrieves all the jobs of a workflow and their associated VM IP addresses from the Pegasus database. It then collects a list of virtual machines owned by a respective user from the Cloud middleware. Using the IP address, it establishes a mapping between the job and the resource configuration of the virtual machine used to execute the job. This information i.e. Cloud-aware provenance is then stored in the provenance store of ReCAP. The flowchart of this mechanism is presented in Figure 8.

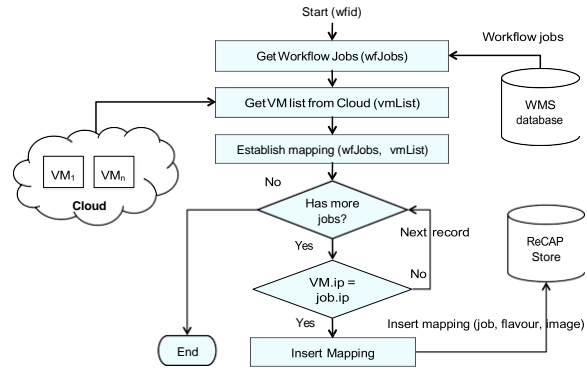

**Figure 8:** Flowchart of creating job-to-Cloud resource mapping using the Static approach

In this flowchart, the variable wfJobs representing a list of jobs of a given workflow is retrieved from the Pegasus database. The variable vmList representing a list of virtual machines in the Cloud infrastructure is collected from the Cloud. A mapping between jobs and VMs is established by match- ing the IP addresses (see in Figure 8). Resource configuration parameters such as flavour and image are obtained once the mapping is established. The *flavour* defines the resource configuration such as RAM, Hard disk and CPUs, and the *image* defines the operating system image used in that particular resource. By combining these two parameters together, one can provision a resource on the Cloud infrastructure. After retrieving these pa- rameters and jobs, the mapping information is then stored in the ReCAP Store (see in Figure 8). This mapping information provides two pieces of important data the: (a) hardware configuration and (b) software configu- ration. (As discussed in ReproRequire Section) These two parameters are important in re-provisioning a similar execution environment.

**Algorithm 1** Job-to-Cloud resource mapping in Static Approach

**Require:** *wfJobs* : Set of jobs in the workflow.
*vmList*: Set of virtual machines in the Cloud infrastructure.

```
 1: procedure JobResourceMapping(wfJobs, vmList)
 2:     cloudResources ← {}
 3:     for all job ∈ wfJobs do
 4:         for all vm ∈ vmList do
 5:             if vm.ip = job.ip then
 6:                 cloudResources[job] ← vm
 7:             end if
 8:         end for
 9:     end for
10:     for all resource ∈ cloudResources do
11:         job ← resource.job
12:         resourceFlavor ← resource.flavor
13:         resourceImage ← resource.image
14:         insertToReCAPStore(job, resourceFlavor, resourceImage)
15:     end for
16: end procedure
```

Algorithm 1 presents the pseudo-code of the Static mapping approach. As discussed previously, this approach cannot work for a dynamic scenario in the Cloud as it establishes the mapping once a workflow has finished and assumes that the machines on which the jobs are executed are still available and accessible. However, in the dynamic situation, Cloud resources will not be available once a job finishes its execution and it will not be possible to retrieve their resource configurations from the Cloud. To overcome this challenging scenario, the following mechanisms *Eager* and *Lazy* are devised which are capable of establishing a job-to-Cloud resource mapping for the Dynamic scenario. These approaches have been discussed in the following sections 7.2 and 7.3.

## *7.2. Eager Approach*

The Eager approach has been devised to establish a job-to-Cloud resource mapping for the dynamic environment on Cloud. In this scenario, a resource may no longer exist on the Cloud when a job has finished. In that case, the information about its configuration cannot be retrieved from the Cloud middleware and the job-to-Cloud resource cannot be established. In this approach, the job-to-Cloud resource mapping is achieved in two phases. In first phase, temporary mapping between the job and Cloud resource is established. It is called temporary mapping because the job is still in running phase and it is possible that it may be scheduled to a different machine in case of job failure. Therefore, this initial mapping is temporary. In the second phase, the final job-to-Cloud resource mapping is established by retrieving job information from the workflow provenance captured by the WMS, which is Pegasus in this research work.

### *7.2.1. Temporary Job-to-Cloud Resource Mapping*

In this phase, the Eager approach monitors the underlying WMS database i.e. Pegasus for the implemented prototype. In Pegasus, along with the host name, its database also maintains the job ID assigned to each job by Condor. The monitoring thread retrieves the condor id assigned to the workflow job and contacts the WMS Wrapper Service (WMS-WS) for information about the job. As the WMS-WS works on top of the underlying workflow man- agement system, therefore it also has an access to the Condor cluster. Upon receiving the request, WMS-WS retrieves job information from the Condor. This information contains the machine IP on which the job is running. Based on this information, the CRM component retrieves the virtual machine in- formation from the Cloud middleware based on the machine IP (as discussed in the Static Approach) and stores this information in the database. This information is treated as temporary because the job is not finished yet and there is a possibility that a job may be re-scheduled to some other machine due to runtime failures. The flowchart of this mechanism is presented in Figure 9.

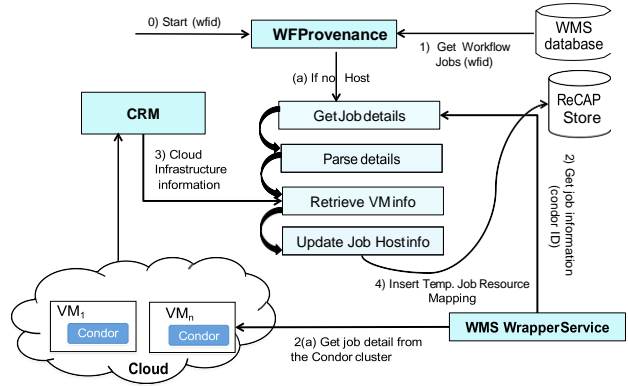

**Figure 9:** Flowchart to create temporary job-to-Cloud resource mapping using the Dynamic approach

### 7.2.2. Final Job-to-Cloud Resource Mapping

This phase starts when the workflow execution is finished. The Provenance Aggregator component starts the job-to-resource mapping process. In doing so, it retrieves the list of workflow jobs from the database and list of vir- tual machines from the Cloud middleware. It starts the mapping between the jobs and the virtual machines based on the IP information, stored in the database, associated with the jobs. In the case of not finding any host information in the database, which is possible in the Dynamic usecase, the Provenance Aggregator retrieves the resource information for that job from the temporary repository that was created in the first phase (as discussed in the above section). Upon finding the Cloud resource information, the Provenance Aggregator component registers this Cloud-aware provenance information in the ReCAP Store. Once the mapping for a job is estab- lished and stored in the database, its corresponding temporary mapping is removed in order to reduce the disk storage overhead. The flowchart of this mechanism is presented in Figure 10. The algorithm of Eager approach is shown in Algorithm 2.

### 7.3. Lazy Approach

The Eager approach is designed to deal with the dynamic Cloud environ- ment. However, it relies upon underlying execution infrastructure such as Condor. To overcome these dependencies, another approach is devised to

**Algorithm 2** Job-to-Cloud resource mapping in Eager Approach

**Require:** *wf Jobs* : Set of jobs in the workflow.

1: *Phase 1: Temporary job-to-Cloud resource mapping*
2: **procedure** JobMonitor(*wfJobs*)
3:    *vmList* ← getVMs
4:   **for all** *job* ∈*wfJobs* **do**
5:      *condorid*← *job.condor*     ▷ each job is assigned unique id
6:      *ip* ← WSClient.getHostInfo*(condorid)*
7:      *vm* ← *vmList*[*ip*]
8:     **if** vm != None & not vmMappingExists(*vm, job*) **then**
9:        resourceFlavor ← vm.flavor
10:       resourceImage ← vm.image
11:       createTempMapping(*job, resourceFlavor, resourceImage*)
12:     **end if**
13:   **end for**
14: **end procedure**
15: *Phase 2: Final job-to-Cloud resource mapping*
16: **procedure** EstablishMapping(*wfJobs*)
17:    *vmList* ← getVMs
18:   **for all** *job* ∈*wfJobs* **do**
19:     **if** job.ip in vmList **then**
20:        vm ← getVM(job.ip)
21:     **else**
22:        vm ← getTempJobMapping(*job*)
23:     **end if**
24:     **if** vm **then**
25:        resourceImage ← vm.image
26:        resourceFlavor ← vm.image
27:        storeJobResourceMapping(*job, resourceFlavor, resourceImage*)
28:        removeTempMapping(*job, resourceFlavor, resourceImage*)
29:     **end if**
30:   **end for**
31: **end procedure**

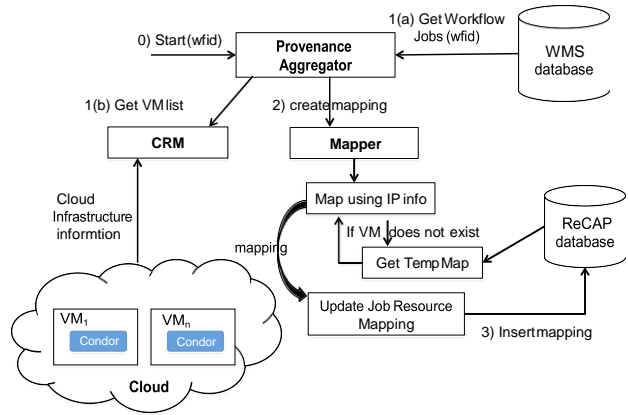

**Figure 10:** Establish the final mapping between Job and VM on Cloud

establish the job-to-Cloud resource mapping for the dynamic Cloud environment. This approach is named *Lazy*. This approach periodically accesses the Cloud and retrieves a list of virtual machines using a monitoring thread that monitors the current status of the available VMs running on the Cloud infrastructure. Each VM along with its creation time is iterated and stored in the ReCAP Store for later use i.e. in the job-to-Cloud resource mapping phase. This is named *Lazy* because it establishes job-to-Cloud resource mapping at the end of a workflow execution. Before, it does not maintain any temporary relation between a job and the virtual machine, unlike the *Eager* approach that maintains a relation between a job and a resource during its phase 1. The algorithm of Lazy approach is presented in Algorithm 3.

This approach periodically monitors the available virtual machines on the Cloud infrastructure and retrieves their metadata information along with their creation time. This information is registered against the VM in a temporary table (see line 10). The database is updated only if new VM information is found on the Cloud (see line 7) against an already existing virtual machine. A new VM is determined mainly by its creation time. Once a job is finished, the mapping will be established using the Jobs host information collected from the Pegasus database and the Cloud resources in- formation from the ReCAP databases. The Cloud resource with the nearest start time from the jobs own starts time and matching IP/hostname is se- lected as a resource for a given job. The advantageous and disadvantageous of this approach are given below.

- **Pros:** This approach may not be efficient in terms of discovery time, but it will work for all scenarios including the static environment be-

**Algorithm 3** Job-to-Cloud resource mapping in Lazy Approach

**Require:** NIL

1:   **procedure** MonitorCloudVirtualLayer(*MONITOR_FLAG*)
2:     **while** MONITOR_FLAG **do**
3:       *vmList* ← getVMs                    ▷ Get a list of VMs
4:       **for all** *vm* ∈ *vmList* **do**
5:         *vm_info* ← *vm.details*
6:         *vm_createtime* ← *vm.creation*
7:         **if** not findVM(*vm_info, vm_creationtime*) **then**    ▷ check for new VM
8:           resourceFlavor ← vm.flavor
9:           resourceImage ← vm.image
10:          insertTempMapping(*vm, resourceFlavor, resourceImage*)
11:        **end if**
12:      **end for**
13:    **end while**
14:  **end procedure**

15: **procedure** EstablishMapping(*wf Jobs*) *It is called when a workflow execution is completed*
16:    **for all** *job* ∈ *wf Jobs* **do**
17:      **if** *job.ip*! = *None* **then**
18:        *ip* ← *job.ip*
19:        *vm* ← getCloudVM*(ip, job.start_time)*   ▷ Get VM for given IP and creation time
20:        **if** vm != None **then**
21:          resourceFlavor ← vm.flavor
22:          resourceImage ← vm.image
23:          insertToReCAPStore(*job, resourceFlavor, resourceImage*)
24:        **end if**
25:      **end if**
26:    **end for**
27: **end procedure**

cause eventually it relies upon host information coming from Pegasus/WMS. It is also essential because during periodic monitoring this approach could not know which job is running one which machine.

- **Cons:** This approach will not be able to determine a mapping between a job and a virtual machine in case where no host information (due to a VM shutdown, job failure or no update in the WMS database) is available in Pegasus. As it does not maintain a temporary resource mapping, it cannot establish job-to-Cloud resource mapping in the absence of host information about a job.

### 7.4. SNoHi Approach

The aforementioned job-to-Cloud resource mapping approaches rely on host information either from the workflow management system's provenance repository or using the underlying infrastructure such as Condor supported by the WMS. However, the proposed mapping approaches will not work if none of these parameters are available. For workflow management systems such as Chimera (Foster et al., 2002) which do not maintain IPs or machine names as part of the job information in their provenance stores, another mapping approach has been devised. This approach has been named *Systems with No Host information* (SNoHi) mapping approach. In this approach, modified job scripts are to be sent that can capture and log machine IP/name information. These jobs can log this information in their output logs if their schema does not permit storing this information in the database. As discussed earlier about the Parser component, for each WMS there will be a dedicated parser plugin. The Parser component will retrieve the job logs from the WMS database and parse the host information. The parsed host information contains the IP and hostname of the resource on which the job was executed. This information is then stored in a temporary table (JobHostTempMap shown in the schema) which maintains only the job-host mapping. At this stage, the Cloud resource configuration information is not stored. Figure 11a illustrates this stage in the SNoHi mapping. Once the job and its host information is available, the SNoHi mapping then can perform the job-to-Cloud resource mapping (as shown in Figure 11b). The Provenance Aggregator component will retrieve the workflow jobs from the WMS database and initiate the mapping. The Mapper will retrieve a list of VMs currently available on the Cloud infrastructure. It will locate the host information from the JobHostTempMap table for workflow jobs and establish the mapping between the job and the Cloud resource. This mapping

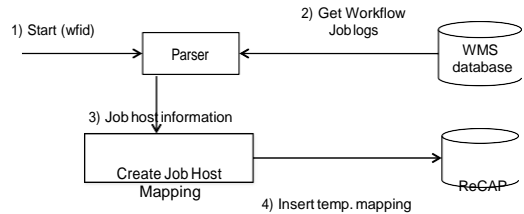

**(a)** Parsing job logs to create temporary job host record

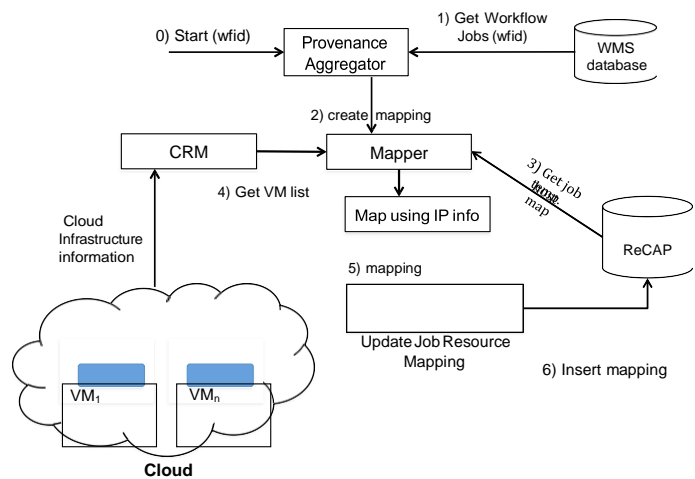

**(b)** Creating final job to Cloud resource mapping information is then stored in the ReCAP database. The SNoHi mapping algorithm is given in Algorithm 4.

**Figure 11:** Creating Job to Cloud resource mapping using the SNoHi approach

---

**Algorithm 4** SNoHi Mapping Algorithm for WMS that does not maintain the machine IP/Name

---

**Require:** *wf Jobs* : jobs of a workflow.

1: **procedure** JobResourceMapping(*wfjobs*)
2:    *jobHostMap* ← getJobHostTempMap*(wfjobs)*
3:    *vmList* ← getVMs
4:    *cloudResource* ← *{}*
5:    **for all** *job* ∈ *jobHostMap* **do**
6:       **for all** *vm* ∈ *vmList* **do**
7:          **if** *vm.ip = job.hostip* **then**
8:             *cloudResources*[*job*] ← vm
9:          **end if**
10:       **end for**
11:    **end for**
12:    **for all** *resource* ∈ *cloudResources* **do**
13:       job ← resource.job
14:       resourceFlavor ← resource.flavor
15:       resourceImage ← resource.image
16:       insertToReCAPStore(*job, resourceFlavor, resourceImage*)
17:    **end for**
18: **end procedure**

---

The algorithm first retrieves the job-host map from the JobHostTempMap table for the given workflow jobs (see line 2). It then retrieves the list of available virtual machines from the Cloud IaaS layer (see line 3). Since the host information has IP of the host, this can be used to establish map-ping between a job and a Cloud resource. Once the job-to-Cloud resource mapping is established, the temporary records are deleted from the Job-HostTempMap table.

## 8. Results and Analysis

In order to verify the proposed approach, ReCAP, it was important to verify that all its components and algorithms were performing according to the theoretical understanding of the design. Since this proposed approach collected

Cloud-aware provenance, it was also important to verify that the captured Cloud resource configurations could indeed affect a job performance and its failure rate, and thus were required in the Cloud environment. In order to evaluate the affect of Cloud configuration on the workflow execution and also to evaluate the proposed mapping approaches in ReCAP, three types of workflows from different scientific domains have been used. These workflows are named; 1) Montage[11] workflow from astronomy domain, 2) ReconAll[12] workflow from neuroscience domain and 3) Wordcount, a sample workflow. The Montage workflow uses the components of Montage, a widely mentioned astronomy application to build mosaics of the sky by stitching together multiple input images (Sakellariou et al., 2010). The Montage workflow used in the experiment contains 35 jobs and required eight input image files. This workflow produces 4 output files including one mosaic image file in JPEG format. The ReconAll workflow, or also known as pipeline, is used in N4U project[13] to reconstruct the given MRI scan image of a subject. This workflow in N4U has only one job that executed recon-all[14] command on the given input neuro-image. The sample workflow i.e. Wordcount is designed for controlled experiments and it exhibits the same characteristics i.e. split and merge jobs found in complex scientific workflows such as Montage. It is composed of four jobs and takes one text file input. The first job (the Split job) took a text file and split it into two files of almost equal length. Later, two jobs (the Analysis jobs) were applied; each of these takes one file as input, and then calculates the number of words in the given file. The fourth job (the Merge job) took the outputs of earlier analysis jobs and calculated the final result i.e. total number of words in both files. Since provenance capturing can also cause performance and size overheads, this paper also presents the results dealt with the overheads (both performance and size) caused by the proposed mapping approaches. The purpose of these experiments was to verify the impact of the devised mapping approaches on the workflow execution. Following subsections provides a detailed analysis of the experiments' results.

*8.1. Resource Configuration impact on Job and Workflow*

Various experiments were performed to analyse the effect of RAM on a job's failure, to analyse the effect of CPU on the job's performance and to analyse the effect of specific resource configurations on the workflow execution performance. This section also discusses the effect of including the CPU MIPS information in the Cloud-aware provenance. The following experiment was designed to evaluate the significance of capturing the virtual machine's RAM parameter in the Cloud-aware provenance. As argued previously in Section 2 this parameter can affect a job's performance as well its failure rate; Fig- ure 12 confirms the effect of the RAM parameter on the job's failure rate. Figure 12 shows that all jobs were successful on all resource configurations until the job's RAM requirement reaches 500 MB. As soon as the jobs mem- ory requirement approaches 500 MB, the job starts failing on the Cloud resource with the m1.tiny configuration because this resource configuration can only provide a maximum of 512 MB of RAM. This memory space is shared among the operating system processes and the job process, conse- quently not enough memory is left for the job. On the other hand, the jobs executed on two other resource configurations i.e. those of m1.small and m1.medium, respectively offering 1024 MB and 2048 MB of RAM respec- tively, were all successful. In this experiment, each job was executed five times with the given memory requirement on each resource configuration. This specific experiment confirms that the RAM can play an important role in job's success rate. This factor is especially important for jobs processing large amounts of data and consequently require more RAM.

---

[11] http://montage.ipac.caltech.edu/
[12] https://surfer.nmr.mgh.harvard.edu/fswiki/recon-all
[13] https://neugrid4you.eu/
[14] https://surfer.nmr.mgh.harvard.edu/fswiki/recon-all

In order to verify the CPU effect on the job performance, a compute intensive job calculating the Fibonacci number was written and executed on different resource configurations (shown in Table 1). The result shown in Figure 13 indicates that a job executed on the *m1.large* resource configuration per- forms better than the job executed on the other resource configurations. The m1.large resource configuration provides more CPU cores and more RAM to the job than the other two resource configurations i.e. m1.small and m1.medium. As can be seen in the figure, the impact of CPU is not evident for the lower ranges of Fibonacci number i.e. 20-30, 30-40. How- ever, as the ranges increases, requiring more computation to calculate the Fibonacci number, the job executed on the improved resource configurations i.e. m1.medium and m1.large performed better. The job took less time on the *m1.large* resource for higher Fibonacci ranges i.e. 40-50 and 50-55.

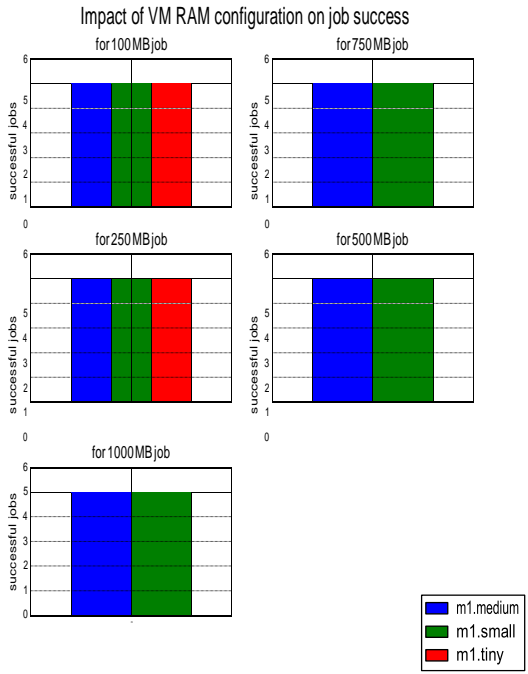

**Figure 12:** A Cloud resource's RAM configuration impact on job success

**Table 1:** Resource Flavours used to execute the compute intensive job

| Flavour   | vCPU | RAM     | Hard Disk |
|-----------|------|---------|-----------|
| m1.small  | 1    | 1024 MB | 10 GB     |
| m1.medium | 2    | 2048 MB | 20 GB     |
| m1.large  | 4    | 4096 MB | 40 GB     |

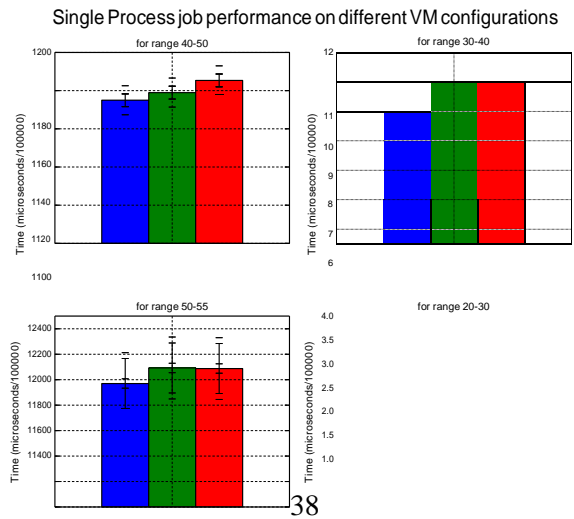

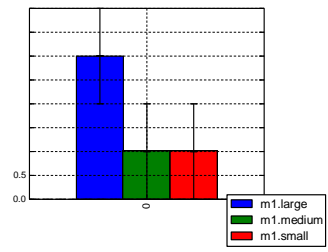

**Figure 13:** Single process job running on different resource configurations

Since Table 1 shows that a few resource configurations offer multiple CPU cores, another compute intensive job was written to calculate the Fibonacci number using parallel programming. This job parallelizes the computing on the available CPU cores on the virtual machine to calculate the Fibonacci number for a given range. The result in Figure 14 shows that a job running on improved resources with parallel programming performs better than a single process job on the same resource. Earlier Figure 13 provide an insight that the performance of a job improves on an improved resource with multiple CPUs. However, the performance of parallel job improves many folds on improved resources (as shown in Figure 14). This means that the number of CPUs in a resource can affect a job's performance that consequently can affect the overall workflow execution performance. This result also confirms that such information about the Cloud resource should be captured as part of the Cloud provenance. Moreover, this factor could play a key role for workflows e.g. the Epigenome workflow in the genomic experiments (Ocana et al., 2011), which is both compute and data-intensive (Pietri et al., 2014), in which analysis performance is also important. Both the results shown in Figure 13 and 14 confirm that it is important to add the CPU information when collecting provenance information because it affects a job execution performance especially for the compute intensive jobs. These results also justify the presence of CPU information in the provenance collected by ReCAP.

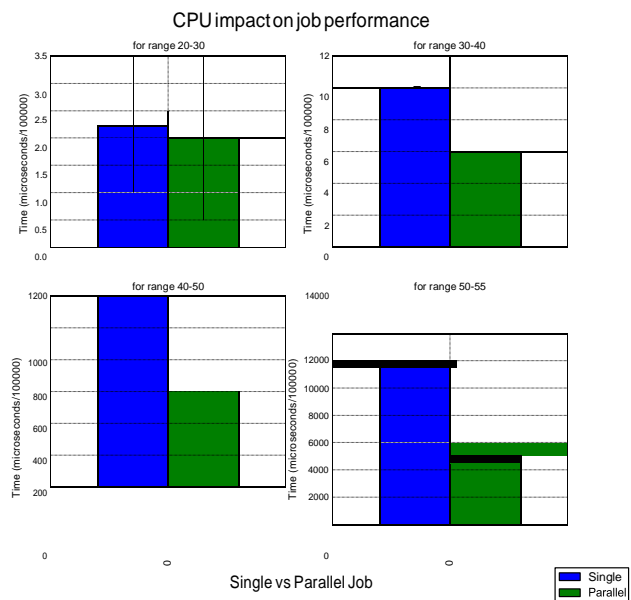

**Figure 14:** Single Process vs Multi-process job running on the *m1.large* resource configuration

This experiment was performed in order to verify the impact of and to highlight the importance of collecting the CPU related data in gathering Cloud-aware provenance information. The motivation behind this experiment came from the findings during the workflow execution on the Condor cluster provisioned on the Cloud infrastructure. It was found that the virtual machines participating in the Condor cluster did not have the same MIPS value (calculated by the Condor benchmarking process) (see Table 2). Moreover, the existing Cloud APIs do not provide access to this information and the resource provisioning request to the Cloud infrastructure also does not include this information. Since this information is very low level and normally determined by benchmarking the resources (as is the case with Condor), this could be one reason that current APIs do not support it yet. On Amazon EC2 page[15] describing the available instance types (or flavours), one can find some indication of CPU speed. However, this information is not accessible through the API. Because of these reasons, this experiment was performed to help in building the support argument for including such information in collecting Cloud-aware provenance information.

**Table 2:** Resource information of used Condor Pools

| Cloud Provider | Arch | OS | RAM | MIPS | KFLOPS |
|---|---|---|---|---|---|
| UWE | x86-64 | Linux | 2002 | 15369 | 1518351 |
|  | x86-64 | Linux | 2002 | 15362 | 1494906 |
| OSDC | x86-64 | Linux | 2003 | 12583 | 1129282 |
|  | x86-64 | Linux | 2003 | 12487 | 1146380 |
|  | x86-64 | Linux | 2003 | 10938 | 1515023 |

Since existing Cloud APIs do not provide this information, a simulated workflow execution was conducted using the WorkflowSim framework Chen and Deelman (2012). This framework allows a user to simulate a datacentre provisioned by describing the virtual resource in terms of the RAM, Hard Disk and CPU MIPS parameters. To mimic the behaviour of different MIPS values in virtual machines a randomly generated MIPS within a small range (12500 with variation of 1500), and a large range variation (10500 with variation of 4500) are assigned to the simulated VMs. Figures 15 and 16 show that execution time is directly affected by a change in the MIPS value. As the MIPS value increases, the execution time decreases and vice versa. In this test, the same Wordcount workflow is simulated on the WorkflowSim but with the given hardcoded execution times.

---

[15]https://aws.amazon.com/ec2/instance-types/

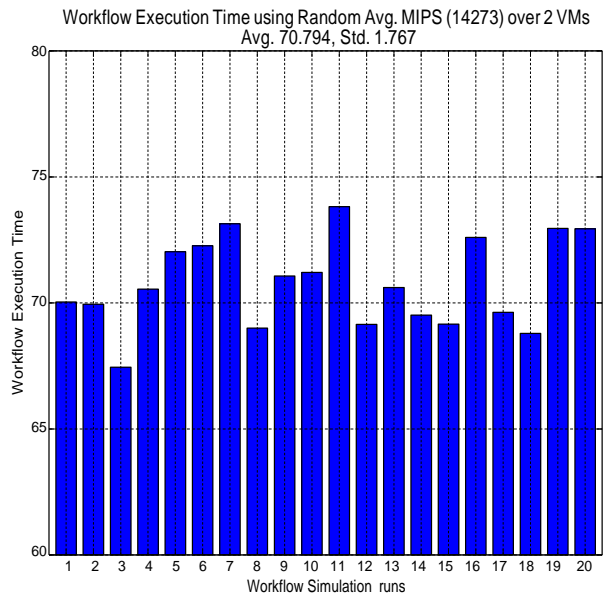

**Figure 15:** Effect of the CPU MIPS on the simulated workflow execution by randomly assigning MIPS from a range to the VMs

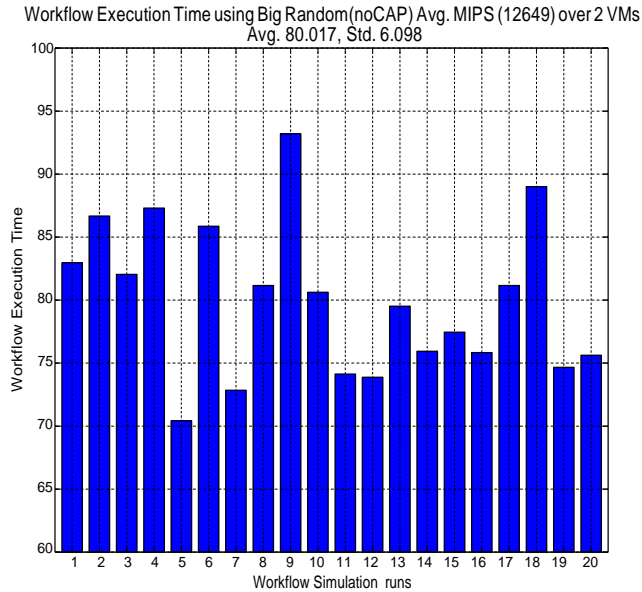

**Figure 16:** The effect of the decrease in the average CPU MIPS on the simulated workflow execution

As the MIPS value increases, the job execution time reduces which results in a better workflow execution performance as is shown in Figure 16. With this result, it can be concluded that the MIPS information along with the col- lected job-to-Cloud resource provenance is very important since it directly affects the job execution time. This information should also be provided while sharing or publishing the results as part of an experimental environ- ment. Seeing the importance of MIPS, it can be argued and proposed that Cloud Providers should also look for ways to offer this as a configuration parameter to a user while provisioning a resource on the Cloud.

Previous experiments evaluated the impact of different resource configurations at the job level. The following set of results aim to show the impact of Cloud resource configuration at the workflow level. In this experiment, the Wordcount workflow was executed at least five times for each resource configuration. Figure 17 shows the average workflow execution time for each configuration type and the small red lines in the figure represent the error bar calculated as standard deviation. Figure 17 shows that the av- erage workflow execution time for the Tiny configuration (i.e. using the m1.tiny flavour type) is higher than the average workflow execution time for the Small and Random configurations. Since the virtual machines pro-visioned with the m1.tiny configuration have less RAM and fewer hard disk resources than the machines with the Small configuration, this is reflected on each job execution time and eventually on the overall workflow execution performance.

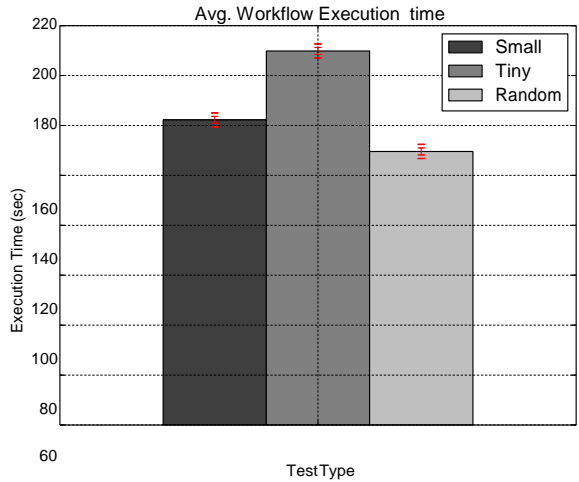

**Figure 17:** Average workflow execution time for each configuration type

The workflow execution time for the Small configuration i.e. m1.small flavour type is better than the Tiny configuration i.e. m1.tiny flavour type. However, the execution time is higher than the execution time obtained using the Random configuration. The reason for the better workflow exe- cution time for the Random configuration can be understood by analysing the provisioned resources. In the Random configuration, the resources were provisioned randomly, which means that a resource could be provisioned with any of the m1.tiny, m1.small, m1.medium or m1.large configurations. In order to further understand the flavour types of these randomly provi- sioned resources, the flavour distribution chart is shown in Figure 18. This figure shows the types of resources which were provisioned in the Random configuration for each workflow execution. From Figure 18, it is clear that all flavour types have been used to provision resources on the Cloud in the Random configuration. Moreover, this figure also shows that more than 50% of the virtual machines were provisioned with the m1.large and m1.medium flavour types, which provide more resources to a virtual machine than the m1.small configuration. Since the jobs have extra resources in terms of the RAM, CPU and Hard disk, this affects their execution times and thus the overall workflow execution time is reduced.

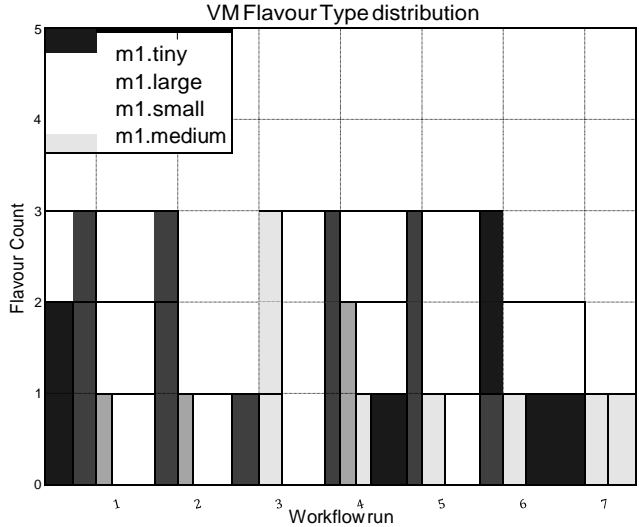

**Figure 18:** Frequency of a flavour type selected randomly during a workflow execution

With these results, it has been established that the factors related to a Cloud resource (as highlighted in Section 2) not only affect a job execution performance and its failure rate, but also can affect a workflow execution performance. The proposed system, ReCAP, is cable of capturing these factors in its Cloud-aware provenance. It can be argued that the scheduler in a workflow management system (WMS) can filter the required resource for a given job in order to avoid job failures, provided the job has announced its required resource configuration. Nonetheless, this information remains useful in acquiring the desired resources from the Cloud instead of acquiring them with random configurations. Moreover, adding these factors in the collected provenance also provides an insight about the types of resources selected by the WMS for job execution. In Figure 17, it was shown that the workflow execution performance improves if the acquired virtual ma- chines have improved resource configurations than the previous execution of the same workflow. However, if the acquired virtual machines do not have better configurations than the original execution, the workflow perfor- mance will deteriorate. Therefore, it can be argued that a user can achieve a similar workflow execution performance by provisioning the similar ex- ecution infrastructure on the Cloud by using the Cloud-aware provenance information.

Moreover, Figures 15 and 16 also illustrate the effect of CPU MIPS on workflow execution time. The simulation results confirm that this parameter should be collected in provenance information. Although the current Cloud APIs do not support the access to this information and also Cloud Providers do not widely support this parameter in resource provisioning pipeline, the result in Figures 16 and 15 show its significance. From the simulation result, it was found that a job's performance was directly related to the MIPS value. As discussed earlier for Figure 13, number of CPUs or CPU speed as MIPs can affect a job performance. This information can be useful for experiments which are both compute and data intensive such as Epigenome workflow in gnomic experiments (Ocana et al., 2011). In the absence of this information, a compute intensive analysis cannot be reproduced correctly because we cannot use this parameter in while requesting a resource.

### 8.2. Infrastructure Re-provisioning

In order to verify that ReCAP can reprovision similar execution infras- tructure on the Cloud using the Cloud-aware provenance, the following ex- periment has been performed. In this experiment, different workflows are executed and their provenance information is captured by ReCAP. Three different workflows have been executed. These three workflows are: (1) Wordcount, (2) Montage and (3) ReconAll. These workflows were submit- ted using ReCAP and their Cloud-aware Provenance was collected. The purpose of this experiment is to evaluate the ability of ReCAP to handle different types of workflows with varying number of jobs and to re-provision resource on the Cloud using the Cloud-aware provenance information. This also evaluates the significance of provisioning similar Cloud infrastructure for workflow re-execution. Following subsections provide the detailed anal- ysis of the results.

- **Using Wordcount Workflow**

The Wordcount workflow was executed using the Pegasus. The original execution of this workflow was assigned an ID 132. Table 3 shows the inter-linked provenance mapping (both the workflow provenance and the Cloud infrastructure information), in the ReCAP database for this workflow. The collected information includes the flavour and image (image name and Im- age id) configuration parameters. The Image id uniquely identifies an OS image hosted on the Cloud and this image contains all the software or li- braries used during the job execution. This workflow is then resubmitted by re-provisioning the resources on the Cloud using the Cloud-aware prove- nance, and its provenance is captured as shown in Table 4. The captured provenance of the reproduced workflow shows that the system was able to re-provision the similar execution resources on the Cloud.

**Table 3:** Cloud-aware Provenance captured for a given workflow

| WfID | nodename | Flavour | RAM (MB) | HD (GB) | vCPU | image name | image Id |
|---|---|---|---|---|---|---|---|
| 132 | uwe-vm3 | 2 | 2048 | 20 | 1 | condorvm-quantal-snapshot | 269cfb39-7882-4067-bf20-b3350a4b1b05 |
| 132 | uwe-vm4 | 2 | 2048 | 20 | 1 | condorvm-quantal-snapshot | 269cfb39-7882-4067-bf20-b3350a4b1b05 |

**Table 4:** Provenance data of the reproduced workflow showing that ReCAP successfully re-provisioned similar resources on the Cloud

| WfID | nodename | Flavour | RAM | HD | vCPU | image name | image Id |
|---|---|---|---|---|---|---|---|
| 134 | uwe-vm4-rep | 2 | 2048 | 20 | 1 | condorvm-quantal-snapshot | 269cfb39-7882-4067-bf20-b3350a4b1b05 |
| 134 | uwe-vm3-rep | 2 | 2048 | 20 | 1 | condorvm-quantal-snapshot | 269cfb39-7882-4067-bf20-b3350a4b1b05 |

In order to measure the execution time of the original workflow and the reproduced workflow on the similar execution infrastructure on the Cloud, the following mechanism has been adopted. The same workflow was executed five times on the Cloud infrastructure. An average execution time was cal- culated for these workflow executions and treated as the average execution time of the original workflow. The ReCAP approach was then used to repro- duce the same workflow execution by re-provisioning the earlier execution infrastructure using the Cloud-aware provenance. The same workflow was re-executed on the re-provisioned resources to measure the execution time of the reproduced workflow. Figures 19 and 20 show the workflow execution times for both the original and reproduced workflows respectively.

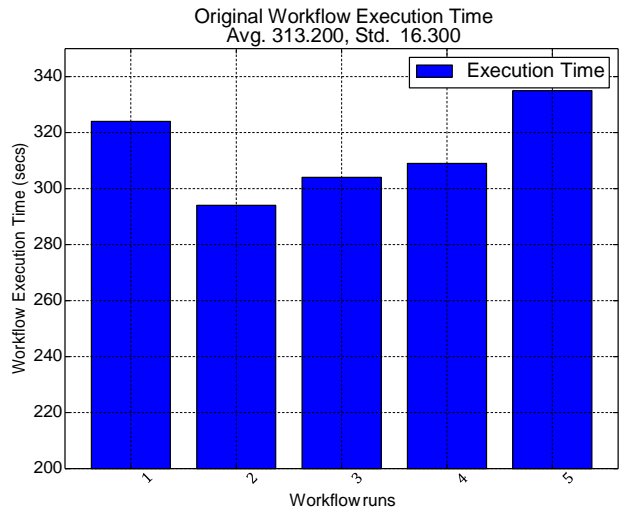

**Figure 19:** Average workflow execution time for the original workflow execution

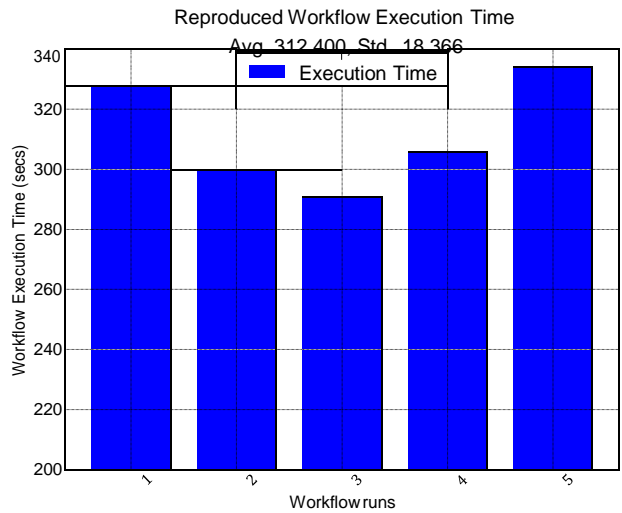

**Figure 20:** Average workflow execution time for the reproduced workflow execution

The average workflow execution time for the original workflow was $313.20 \pm 16.30$ seconds and the average workflow execution time for the reproduced workflow on the similar execution infrastructure was $312.400 \pm 18.366$ seconds. This shows that the execution time for the reproduced workflow is almost similar to the original workflow execution time. This result shows that we can expect similar workflow execution performance provided similar execution infrastructure was provisioned.

- **Using Montage Workflow**

The original execution of this workflow was assigned an ID 533 and the reproduced workflow was assigned an ID 534. Following tables 5 and 6 show the Cloud infrastructure provisioned for the execution of workflows 533 and 534 respectively. The workflow execution times for the original workflow 533 and reproduced workflow 534 were 285 and 281 seconds respectively. Table 6 shows that the Montage workflow execution was reproduced with a similar execution performance on the same Cloud resources provisioned by ReCAP.

**Table 5:** Infrastructure detail captured for the Montage workflow (wfID=533) using ReCAP

| WfID | nodename | Flavour | RAM | HD | vCPU | image name | image Id |
|---|---|---|---|---|---|---|---|
| 533 | uwe-vm3 | 2 | 2048 | 20 | 1 | wf_peg_repeat | f102960c-557c-4253-8277-2df5ffe3c169 |
| 533 | mon1 | 2 | 2048 | 20 | 1 | montage-condor-setup | 2d9787e4-b0e9-4802-bf4f-4a0c868bb11a |

**Table 6:** Infrastructure detail captured for the reproduced Montage workflow (wfID=534) using ReCAP

| WfID | nodename | Flavour | RAM | HD | vCPU | image name | image Id |
|---|---|---|---|---|---|---|---|
| 534 | mon1-rep | 2 | 2048 | 20 | 1 | montage-condor-setup | 2d9787e4-b0e9-4802-bf4f-4a0c868bb11a |
| 534 | uwe-vm3-rep | 2 | 2048 | 20 | 1 | wf_peg_repeat | f102960c-557c-4253-8277-2df5ffe3c169 |

- **Using ReconAll Workflow**

The ReconAll workflow consists of one job script only and requires one input file. Its execution requires one virtual machine on the Cloud. The original execution of this workflow was assigned ID 545 and its captured Cloud-aware provenance using ReCAP is shown in Table 7. The same workflow was re-executed using ReCAP and its provenance information is recaptured. The reproduced workflow execution was assigned an ID 547. The recaptured provenance of the reproduced workflow 547 is show in Table 8. These tables show that ReCAP was able to capture the Cloud infrastructure information in the original execution and was also able to re-provision the resource with the same configurations on the Cloud. The execution time of the original workflow was 32599.16 ± 158.73 seconds and the execution time of the reproduced workflow was 32620.0 ± 147.78 seconds.

**Table 7:** Infrastructure detail captured for the ReconAll workflow (wfID=545) using ReCAP

| WfID | nodename | Flavour | RAM | HD | vCPU | image name | image Id |
|---|---|---|---|---|---|---|---|
| 545 | freesurf | 2 | 2048 | 20 | 1 | freesurf-condor | 2ee3c500-61b5-4592-8d54-e572536b5df1 |

**Table 8:** Infrastructure detail captured for the ReconAll reproduced workflow (wfID=547) using ReCAP

| WfID | nodename | Flavour | RAM | HD | vCPU | image name | image Id |
|---|---|---|---|---|---|---|---|
| 547 | freesurf-rep | 2 | 2048 | 20 | 1 | freesurf-condor | 2ee3c500-61b5-4592-8d54-e572536b5df1 |

In order to see the effect of resource configurations on the job execution performance, and subsequently on the workflow execution time, the same workflow was re-executed on a resource with different configuration (see Table 9). This time the resource was provisioned manually but its provenance was captured through ReCAP. The workflow execution time in this case was 32278.5 seconds. From this result, it must be noted that the execution time decreases on a resource with better resource configurations i.e. more RAM and CPUs. This result verifies the original argument that resource configurations can affect a job execution and thus overall workflow execution performance. This is the reason why capturing resource configurations for workflows execution on the Cloud is essential.

**Table 9:** Infrastructure details captured for the workflow (wfID=554) using ReCAP

| WfID | nodename | Flavour | RAM | HD | vCPU | image name | image Id |
|---|---|---|---|---|---|---|---|
| 554 | newfree | 3 | 4096 | 40 | 2 | freesurf-condor | 2ee3c500-61b5-4592-8d54-e572536b5df1 |

### 8.3. Mapping approaches Overhead

In order to evaluate the impact of the proposed provenance capture i.e. the mapping approaches on the workflow execution performance, a modified Wordcount workflow was executed with different provenance capturing approaches. This workflow also has the same four jobs but the jobs in this workflow sleep, to mimic the job processing time instead of actually processing the data, for a given time period, that was passed as an argument to them. This was done in order to measure the impact of only provenance mapping approaches on the workflow execution performance by eliminating the effect of the data transfer time on the workflow execution time. The *preprocess* job sleeps for 120 seconds, the *analysis1* job sleeps for 120 seconds, the *analysis2* job sleeps for 60 seconds and the *merge* job sleeps for 60 seconds. The workflow was executed with three different provenance mapping approaches proposed in this work. In order to evaluate both static and dynamic mapping approaches, both the Static mapping approach and the Eager mapping approach was applied. The third mapping approach i.e. the SNoHi Mapping approach (discussed in Section 7.4) was also tested to analyse its impact on the workflow execution performance. For the third mapping approach, the workflow jobs were further modified to collect the machine information during the job execution. The workflow was executed 10 times for each mapping approach. The workflow execution performance in the presence of these mapping approaches was compared against the work- flow execution performance in the absence of these mapping approaches.

The results shown in Figure 21 confirms the theoretical understanding of the proposed mapping approaches (as discussed in Section 7) because it complies with the intended outcomes. As it can be seen there is no great difference in the workflow execution times in the presence of the Static and Eager mapping approaches. The average workflow execution time in the absence of any provenance approach is $434.67 \pm 6.52$ seconds. The average workflow execution time with the Static Mapping approach is $434.9 \pm 4.29$ seconds and the average workflow execution under the Eager approach is $434.78 \pm 4.68$ seconds. The main reason for these mapping approaches not having a major impact on the workflow execution time is because the proposed mapping approaches work outside the virtual machines, thus they don't interfere with the job execution. However, this is not the case with the SNoHi mapping approach in which the provenance information was captured within the job itself. Due to this, there is a small increase in the workflow execution time for the SNoHi mapping approach. The average workflow execution time for the SNoHi mapping approach is $435.56 \pm 4.83$ seconds. It was expected that the workflow execution time under the SNoHi mapping approach would increase because an additional provenance collection process runs within the job, which increases the job execution time, hence it affects the overall workflow execution time. However, there is no significance increase in the workflow execution time observed with the SNoHi mapping approach. This is because the provenance collection process during the job execution took only $0.0418 \pm 0.0017$ seconds (on average).

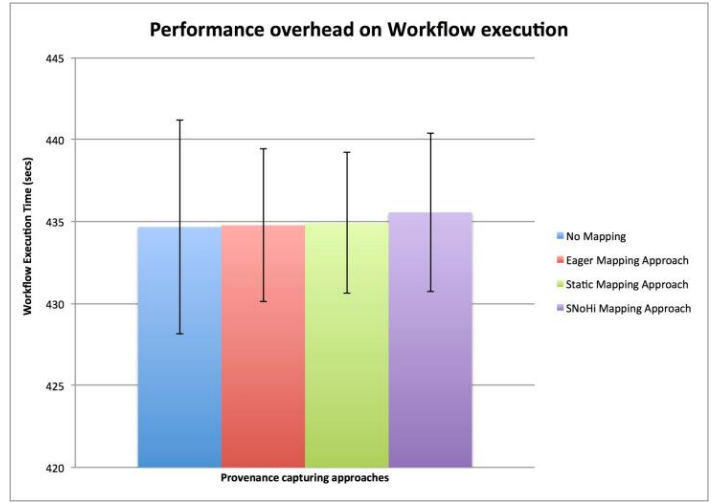

**Figure 21:** Effect of different provenance collection approaches on the workflow execution performance

## 9. Conclusions and Future Work

The dynamic nature of the Cloud makes provenance capturing of workflow(s) with the underlying execution environment(s) and their reproducibility a difficult challenge. In this regard, a list of workflow reproducibility requirements has been presented in this paper after analysing the literature and workflow execution scenario on the Cloud infrastructure. The proposed ReCAP's framework can augment the existing workflow provenance with the Cloud infrastructure information to generate the Cloud-aware provenance. In this paper, two resource usage scenarios i.e. Static and Dynamic on the Cloud have been identified. Based on the identified Cloud usage scenarios, different mapping approaches have been proposed. These mapping approaches in ReCAP collected the workflow provenance from the underlying workflow management system and the Cloud provenance from the Cloud infrastructure, and then linked them together. This mapping information was then used to re-provision resources on the Cloud and to re-produce a workflow execution. Once an execution infrastructure was re-provisioned, the workflow was re-submitted for execution and its provenance was again captured. The provenance of the reproduced workflow was compared against the original workflow in order to determine the reproducibility. The results (see Tables 3, 5 and 7) show that the proposed approaches can capture the Cloud-aware provenance (CAP) by capturing the information related to Cloud infrastructure (virtual machines) used during a workflow execu- tion. It can then re-provision a similar execution infrastructure with same resource configurations on the Cloud using CAP to reproduce a workflow execution (see Tables 4 and 8). Figure 19 shows that the workflow execution time remains the same for reproduced workflow because similar execution infrastructure was provisioned using the Cloud-aware provenance. In order to evaluate the proposed work, three different workflows 1) *Montage* from astronomy domain 2) *ReconAll* from neuroscience domain and 3) *Wordcount* - a simulated workflow for controlled analysis - were executed using ReCAP and their provenance was captured for analysis. The results were found to be consistent for all these three workflows. By using different workflows from different scientific domains, the applicability of the proposed approach Re-CAP in different domains was also demonstrated. Furthermore, this paper also presents the impact of the devised mapping approaches on the workflow execution time. The result in Figure 21 shows that the presented mapping approaches do not significantly affect the workflow execution time because they work outside the virtual machine.

As discussed in the workflow reproducibility requirements (see Section 4), keeping track of workflow versions and their associated files has been high-lighted in the literature as an important point for workflow reproducibility. At present, the ReCAPs database schema does not support multiple ver- sions of a workflow and their evolution. As a future work, the ReCAP system can be integrated with systems such as CRISTAL or VisTrails that can keep track of such evolutions to support multiple versions of a work- flow. At present, ReCAP relies upon the OS image stored on the Cloud as the basis for providing all the required software and the operating sys- tem stack for the virtual resource. However, this limitation can be avoided with a mechanism that can configure a base virtual machine to the required level by installing and configuring all the required software. One such ap- proach (Klinginsmith et al., 2011) helps in re-installing and reconfiguring a software environment on top of VMs. In future, the proposed work can be extended by integrating it with such approaches to provide a fully automated mechanism to re-provision similar resources with a completely configurable software stack on the Cloud for workflow re-execution. Besides these, the Cloud-aware provenance information collected by ReCAP can also be used in workflow planning and scheduling algorithms. Moreover, the captured

Cloud-aware provenance can be used to intelligently provision the execution resources on-demand and to schedule jobs onto them in order to provide efficient resource utilization. Furthermore, ReCAP has been evaluated with different workflows with varying number of jobs, which were executed over a moderate number of Cloud resources due to the limited resources avail- able in the project. ReCAPs scalability evaluation over a very large pool of Cloud resources is also an intended future work.

## 10. Acknowledgements

The authors would like to acknowledge the support of the European Union in funding this work via the neuGRID4You (N4U) project (grant agreement n. 283562, 2011-2015).